\documentclass[twocolumn]{aastex631}

\usepackage{graphicx}	
\usepackage{amsmath}	
\usepackage{booktabs}
\usepackage{xcolor}

\begin{document}
\title[NGC 7314]{NGC 7314: X-ray Study of the Evolving Accretion Properties}
\author[0000-0001-6770-8351]{Debjit Chatterjee}
\affiliation{Institute of Astronomy, National Tsing Hua University, Hsinchu 300044, Taiwan}
\affiliation{Indian Institute of Astrophysics, II Block Koramangala, Bangalore 560034, India}

\author{Arghajit Jana}
\affiliation{Institute of Astronomy, National Tsing Hua University, Hsinchu 300044, Taiwan}
\affiliation{N{\'u}cleo de Astronom{\'i}a de la Facultad de Ingenier{\'i}a, Universidad Diego Portales, Av. Ej{\'e}rcito Libertador 441, Santiago, Chile}

\author{A. Mangalam}
\affiliation{Indian Institute of Astrophysics, II Block Koramangala, Bangalore 560034, India}

\author{Hsiang-Kuang Chang}
\affiliation{Institute of Astronomy, National Tsing Hua University, Hsinchu 300044, Taiwan}
\affiliation{Department of Physics, National Tsing Hua University, Hsinchu 300044, Taiwan}

\begin{abstract}

We present a comprehensive analysis of the timing and spectral properties of NGC 7314, a Seyfert 1.9 galaxy, using X-ray observations from {\it XMM-Newton}, {\it NuSTAR}, and {\it RXTE}/PCA. The timing analysis reveals significant variability across different energy bands, with fractional variability (F$_{\rm var}$) values consistent with previous studies. The highly variable soft photons and comparatively less variable high energy photons imply different origins of these two types. The soft energy photons come from a hot corona near the center, while the high-energy photons are produced by inverse Compton scattering of these primary X-ray photons in a hot plasma away from the central region. The spectral analysis employs various models to characterize the emission components. The results indicate the presence of a soft energy bump, Fe K$\alpha$ line emission, and a prominent reflection component. The long-term {\it RXTE}/PCA data analysis reveals temporal variations in the photon index ($\Gamma$) and power-law flux, suggesting evolving emission properties over time. The signature of both broad and narrow Fe~K$\alpha$ emission line features suggested the broad, variable one coming from the accretion disk ($\sim10^{-5}$~pc), while the non-evolving narrow line can not be well constrained. The absorption feature could originate in a highly ionized region, possibly closer to the broad-line region (BLR). The evolution of the inner accretion properties indicates that NGC 7314 could be a potential changing-state active galactic nuclei.
\end{abstract}



\section{Introduction}
Active galactic nuclei (AGN) are considered to be powered by accretion onto a supermassive black hole \citep{Rees1984} at the galactic center of their host galaxies. The infalling matter with angular momentum is expected to form an accretion disk around the central compact object. Accretion into a supermassive black hole leads by converting the gravitational potential energy to the observed radiation, spanning the entire electromagnetic band (from radio to $\gamma$-rays). A large part of this energy dissipates as X-ray emission very close to the central black hole. The classification of AGNs depends on various parameters, including orientation \citep{Antonucci1993,Urry&Padovani1995,Netzer2015}, accretion rate \citep{Heckman&Best2014}, the presence (or
absence) of strong jets \citep{Padovani2016}. Other factors, such as the host galaxy and its environment, also play a role. The AGNs in the Seyfert class are commonly categorized into two distinct types based on their optical emission line characteristics- Seyfert 1 and Seyfert 2. \citet{Osterbrock1981} introduced a more detailed classification for Seyfert galaxies based on the strength of the broad H$\alpha$ and H$\beta$ emission lines compared to the narrow lines. This classification includes intermediate types, numbered from 1.2 to 1.9, depending on how strong these broad lines are. For example, in Seyfert 1.9 galaxies, the broad H$\alpha$ line is visible (weak), but the broad H$\beta$ line is not. These different classifications are explained by the orientation of the viewing angle with respect to the circumnuclear molecular torus. The Seyfert 1 subgroup is classified for their face-on view with respect to the observer. They give a relatively unobscured view of the central engine and are considered to be fruitful in an observational sense. On the other hand, Seyfert 2 galaxies are observed at high inclination angle\citep{Antonucci1993,RamosAlmeida2017}. This geometry causes the central engine to be entirely blocked by the dusty torus surrounding it. 

The observed X-ray emission from an AGN is considered to be mainly due to the thermal Comptonization of the soft optical and ultra-violet photons from the disk\citep{Haardt&Maraschi1993}. These soft photons get inverse Comptonized in a cloud of hot electrons produced by the inner part of the accretion disk, called corona \citep{Haardt&Maraschi1991,Haardt&Maraschi1993}. A power-law with an exponential cut-off at high energy can represent this radiation (also called primary emission). The primary continuum can be reprocessed by the dusty torus or/and different parts in the accretion disk, producing a `reflection hump' around $20-30$ keV. Reflection spectrum from distant matter generates a neutral Fe K$\alpha$ line emission at $6.4$ keV \citep{George&Fabian1991,Matt1991,Mushotzky1993}. Reflection close to the SMBH gives a broadened fluorescent line (width $\sim$ 1 keV) due to the gravitational and Doppler effect \citep{Fabian1989,Fabian2000}. Some Seyfert 1 galaxies exhibit an excess of soft X-rays below 2 keV \citep{Singh1985}, a feature that remains a topic of debate regarding its generation mechanism. Studies suggest that the soft excess is due to Comptonization in a warm or hot corona or to relativistic reflection from the accretion disk \citep{Walter&Fink1993,Piconcelli2005,Done2012,Nandi2023}.

NGC 7314 ($z\sim0.004763$ \cite{Matheson1996}) is a spiral galaxy classified as SAB(rs)bc, with an AGN at its center. The O~\textsc{i} emission line at $\lambda=8446~A$ (excited by Bowen fluorescence mechanism) from the spectro-photometry study NGC 7314 leads to categorized this as type-I Seyfert galaxy\citep{Morris1985}. However, in a later study of the spectrum of the nuclear region revealed a broad component of H$\alpha$, classified it as a Seyfert 1.9 type \citep{Hughes2003}. 
In a recent study of 20 Seyfert 1 galaxies with RXTE observations, sample NGC 7314 is categorized as broad line Seyfert 1 (BLS1) \citep{Weng2020}. The mass of the central black hole is $\sim$5$\times$10$^6$ $M_\odot$ \citep{Schulz1994}. NGC 7314 shows rapid variability in X-ray \citep{Turner1987,Yaqoob1996,Yaqoob2003}. The narrow and broad components of the Fe K line exhibit different patterns of variability in response to changes in the illuminating continuum \citep{Yaqoob2003}. The Fe K lines at 6-7 keV energy band lag both the lower and higher energy bands and are consistent with relativistically broadened iron K$\alpha$ line\citep{Zoghbi2013}. A recent multi-wavelength study of the nuclear and circumnuclear emission of NGC 7314 revealed that in the observed optical spectrum, the emission from a Seyfert nucleus is evident, displaying broad components within the H~$\alpha$ and H~$\beta$ emission lines \citep{daSilva2023}. The study concluded the presence of a type 1 AGN that showcases a spectrum abundant in coronal emission lines. The spatial characteristics revealed an ionization cone to the west of the nucleus, while the east cone's visibility is compromised due to dust obstruction. Analysis of X-ray data indicates fluctuations in flux; however, they have not noticed any variations in the line of sight's column density. The study suggested that the variability could potentially originate from AGN's accretion rate fluctuations.

This paper aims to analyze the timing and spectral properties of NGC 7314, a Seyfert 1.9 galaxy, using X-ray observations from the {\it XMM-Newton}, {\it NuSTAR}, and {\it RXTE}/PCA satellites. The study focuses on understanding the variability in different energy bands, the origins of the soft and high-energy photons, and the evolution of the inner accretion properties. The analysis reveals significant variability, the presence of a soft energy bump, Fe K$\alpha$ line emission, and a prominent reflection component. The paper further explores the temporal variations in the photon index and power-law flux. The paper is organized in the following way. In Section \S2, we describe the observation and data analysis processes. The results obtained from our timing and spectral analysis are presented in Section \S3. In Section \S4, we discuss our findings. We assume a cosmological model with $\Omega_\Lambda=0.7$, $\Omega_M=0.3$, and $H_0=
70$ km~s$^{-1}$~Mpc$^{-1}$.

\begin{table*}
	\centering
	\caption{NGC 7314 observation log}
	\label{tab:table1}
	\begin{tabular}{lccr} 
		\hline
		Satellite/ & Obs Id & Date & Exposure\\
		Instrument &        & (dd-mm-yyyy) & ($\sim$ks)\\
         Col. 1 & Col. 2 & Col. 3 & Col. 4 \\
		\hline
        {\it RXTE} & 85 Obs & 01-01-1999 -- 16-07-2000 &--\\ 
                   &  7 Obs & 19-07-2002 -- 22-07-2002 &--\\
		{\it NuSTAR} & 60201031002 & 13-05-2016 & 200\\
        {\it XMM-Newton} & 0111790101 & 02-05-2001 & 44\\
                   & 0725200101 & 17-05-2013 & 140\\
                   & 0725200301 & 28-11-2013 & 132\\
                   & 0790650101 & 14-05-2016 & 65\\
		\hline
	\end{tabular}

 \noindent{The detailed list of our studied observations. The observatories/satellites' names are given in Col. 1. Col. 2 represents the observation Ids. of the respective satellites. For {\it RXTE}, the total number of observations is given. Col. 3 shows the observation date in dd/mm/yyyy format. For {\it RXTE} the range of the date is given. Col. 4 represents the total exposure time in ks. Note: The instruments used for the different satellites are -- {\it RXTE} PCA, {\it NuSTAR} FPMA and FPMB, {\it XMM-Newton} EPIC-pn.}
 
\end{table*}

\section{Observation and Data Reduction}
We used publicly available archival data of {\it RXTE}/PCA  and {\it NuSTAR} from HEASARC\footnote{https://heasarc.gsfc.nasa.gov/cgi-bin/W3Browse/w3browse.pl}. The {\it XMM-Newton} data were downloaded from XMM-Newton Science Archive\footnote{http://nxsa.esac.esa.int/nxsa-web/\#search}. The summary of the observations taken for the study is given in Table~\ref{tab:table1}.

\subsection{RXTE}

We used a total of 92 archival data of the proportional counter array (PCA) onboard {\it RXTE} \citep{Bradt1993} from January 01, 1999 (MJD=51179.74) to July 16, 2000 (MJD=51741.09) and from July 19, 2002 (MJD=52474.22) to July 22, 2002 (MJD=52477.71). We followed the standard procedure to extract the PCU2 spectra described in 'The ABC of XTE'\footnote{https://heasarc.gsfc.nasa.gov/docs/xte/abc/front\_page.html}. For the spectral analysis, we use {\tt Standard2f} mode data with 16-s time resolution, which has 128 energy channels. Spectra are extracted only from PCU2. We generated a background fits file using the {\tt PCABACKEST} tool and the 'bright' background model appropriate for our observation periods. A good time interval (GTI) file is created using the FTOOLS task {\tt maketime} to include only periods when the instrument operates under optimal conditions. The {\tt saextrct} tool was used to extract the source spectra and background spectra using the GTI file. To improve the signal-to-noise ratio, we rebinned the spectra using the {\tt rbnpha} tool, combining adjacent energy channels to ensure each bin had a minimum number of counts. The spectra were rebinned to have at least 5 counts/bin to obtain valid $\chi^2$ statistics. We then generated the response matrix and effective area files using the {\tt pcarsp} tool to account for the instrumental response.

\subsection{NuSTAR}
{\it NuSTAR} observed NGC 7314 on May 13, 2016. {\it NuSTAR} consists of two identical focal plane modules- FPMA and FPMB \citep{Harrison2013}. The {\it NuSTAR} raw data was reprocessed using the {\it NuSTAR Data Analysis Software} ({\tt NuSTARDAS version 2.1.2}). Calibrated and cleaned event files were generated by {\tt nupipeline} task. We used 20200912 version of calibration files from {\it NuSTAR} calibration database\footnote{http://heasarc.gsfc.nasa.gov/FTP/caldb/data/nustar/fpm/}.  We used 60'' circular regions to extract both the source and background spectra. The background region was selected far away from the source region of the same chip. The light curves and spectra were produced from the cleaned science mode event files through {\tt nuproducts} task. Light curves were extracted with 300 sec time binning. The light curves from two modules were combined with {\tt lcmath} task. For the variability study, we produced $3-10$ keV (soft band), $10-78$ keV, and $3-78$ keV light curves. We rebinned the $3-78$ keV spectra with 20 counts/bin using {\tt grppha} task.

\subsection{XMM-Newton}
NGC 7314 was first observed on May 02, 2001 by {\it XMM-Newton}. Among the two observations on that day, we used only ObsId. 0111790101 for its comparatively high exposure. We also used two observations 0725200101 and 0725200301, from May 17, 2013, and November 28, 2013. For a simultaneous study with the {\it NuSTAR} observation, we used ObsId. 0790650101 from May 14, 2016 with 65~ks exposure. The observation data files (ODFs) from the European Photon Imaging Camera (EPIC) on the detector were processed using the Science Analysis System ({\tt SAS}; \citealt{Gabriel2004}) version 20.0.0). We followed standard procedures\footnote{https://www.cosmos.esa.int/web/xmm-newton/sas-threads} to obtain calibrated and concatenated event lists, by filtering them for periods of high background flaring activity, and by extracting the light curves and spectra. The source events were extracted using a circular region, with a radius of 36'' arcsec, centered on the target, and the background events were extracted from a circular region, with a radius of 40'' arcsec, on the same chip far from the source. We verified that the photon pile-up is negligible in the filtered event list with the task {\tt epatplot}. After that, the response matrix files (RMFs) and ancillary response files (ARFs) were generated, and the spectra were re-binned in order to include a minimum of 25 counts in each background-subtracted spectral channel and, also, in order to not oversample the intrinsic energy resolution by a factor larger than 3. We have also extracted $0.2-3$ keV, $3-10$ keV, and $0.2-10$ keV light curves for the four {\it XMM-Newton} observations with 300~sec binning to study the variability. We followed the procedure in the webpage\footnote{https://www.cosmos.esa.int/web/xmm-newton/sas-thread-timing} for extracting light curves.

\section{Results}
\subsection{Timing analysis}
In X-ray binary studies, it is customary to examine timing properties using Power Spectral Densities (PSDs) averaged over multiple light curves to reduce noise \citep{vanderKlis1995}. However, AGN studies often rely on single light curves due to limited data, which can be misleading, as fluctuations in variance may simply reflect the stochastic nature of the process rather than genuine physical changes \cite{PapadakisLawrence1993,Uttley2002}. The excess variance statistic ($F_{\rm var}$) can be employed to quantify variability in AGNs, even with limited observational data. The excess variance can reveal valuable information despite the difficulties of robustly estimating variability amplitudes from short observations. For instance, it has been shown that the variability amplitude in Seyfert 1 galaxies is inversely correlated with the source luminosity \citep{Nandra1997, Leighly1999, MarkowitzEdelson2001}. Additionally, differences in the normalized excess variance between energy bands can indicate energy-dependent PSDs or independently varying spectral components, further enriching our understanding of AGN variability.

To characterize the extent of variability in our data, we utilized the normalized excess variance ($F_{\rm var}$) as a measure. The $F_{\rm var}$ parameter, introduced by \cite{Edelson2002} and further discussed by \cite{Vaughan2003}, enables us to quantify the intrinsic variations of the source while mitigating the impact of measurement errors.

In accordance with the methodology outlined in \cite{Vaughan2003}, we define the $F_{\rm var}$ as follows:

\begin{equation}
F_{\rm var} = \sqrt{\frac{S^{2} - \bar{\sigma}^{2}_{\rm err}}{\bar{x}^2}}
\end{equation}

In this equation, $S^2$ denotes the sample variance, $\bar{x}$ represents the arithmetic mean of the data points $x_i$, and $\bar{\sigma}^{2}_{\rm err}$ is the average of the squared measurement errors.

The values of $S^2$ and ${\bar{\sigma}^{2}_{\rm err}}$ are computed as follows:

\begin{equation}
S^2 = \frac{1}{N-1} \sum_{i=1}^{N} (x_i - \bar{x})^2
\end{equation}

\begin{equation}
\bar{\sigma}^{2}_{\rm err} = \frac{1}{N} \sum_{i=1}^{N} \sigma^{2}_{\rm{err,i}}
\end{equation}

To estimate the uncertainty associated with the $F_{\rm var}$ value, we use the following formula:

\begin{equation}
\text{err}(F_{\rm var}) = \sqrt{\left(\sqrt{\frac{1}{2N}} \frac{\bar{\sigma}^{2}_{\rm err}}{\bar{x}^2} F_{\rm var}\right)^2 + \left(\sqrt{\frac{\bar{\sigma}^{2}_{\rm err}}{N}} \frac{1}{\bar{x}}\right)^2}
\end{equation}

This equation takes into account the contribution of both the variance in measurement errors and the uncertainties related to the mean and $F_{\rm var}$ value.

\begin{figure}
	\includegraphics[width=8.5cm]{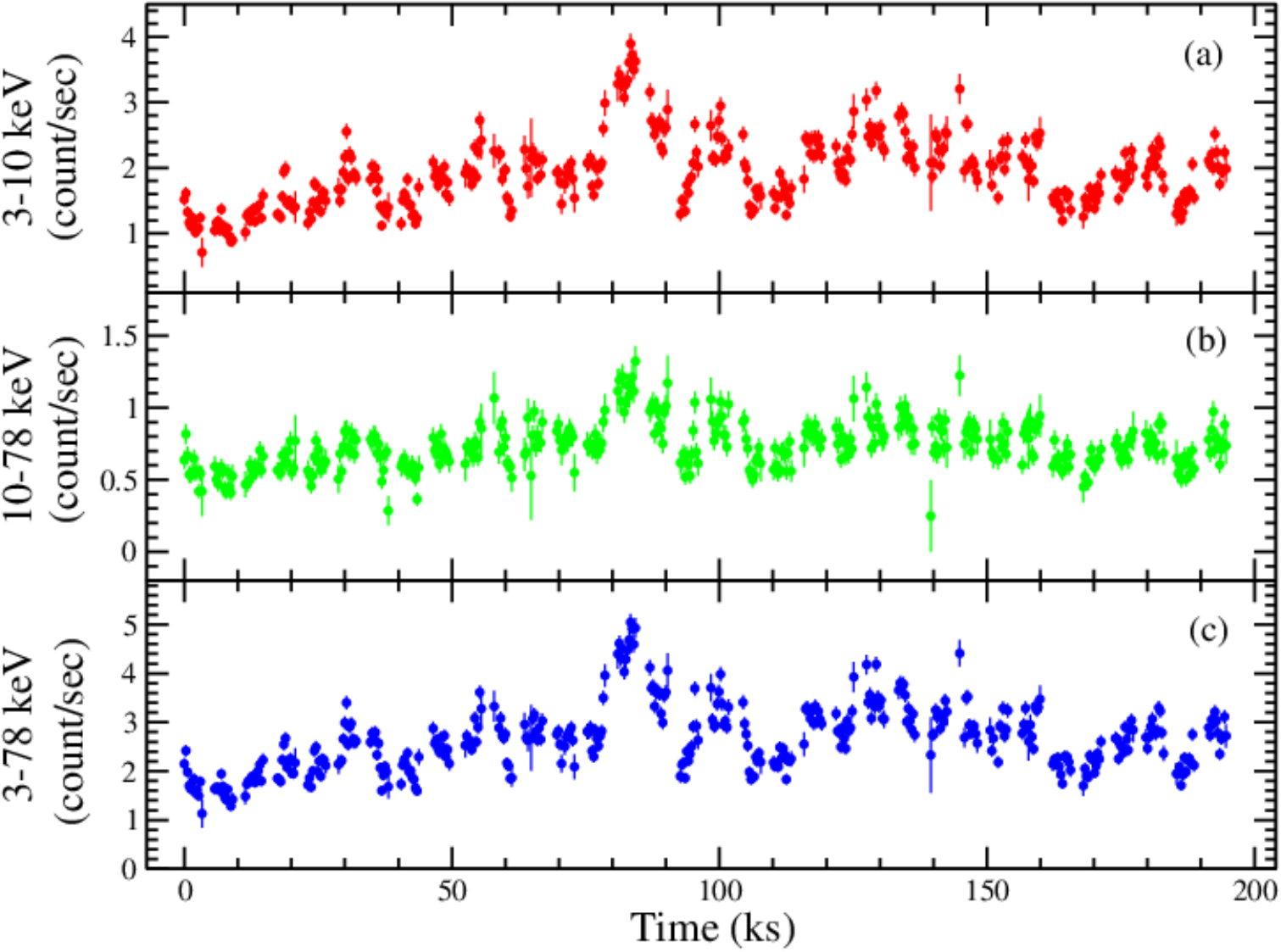}
    \caption{300 sec time binned light curves of combined {\it NuSTAR} FPMA and FPMB.
    The three panels represent (a) $3-10$ keV (soft; \textcolor{red}{red}), (b) $10-78$ keV (hard; \textcolor{green}{green}), and (c) $3-78$ keV (total; \textcolor{blue}{blue}) light curves respectively.}
    \label{fig:lcurv}
\end{figure}

\begin{figure*}
	\includegraphics[width=16cm]{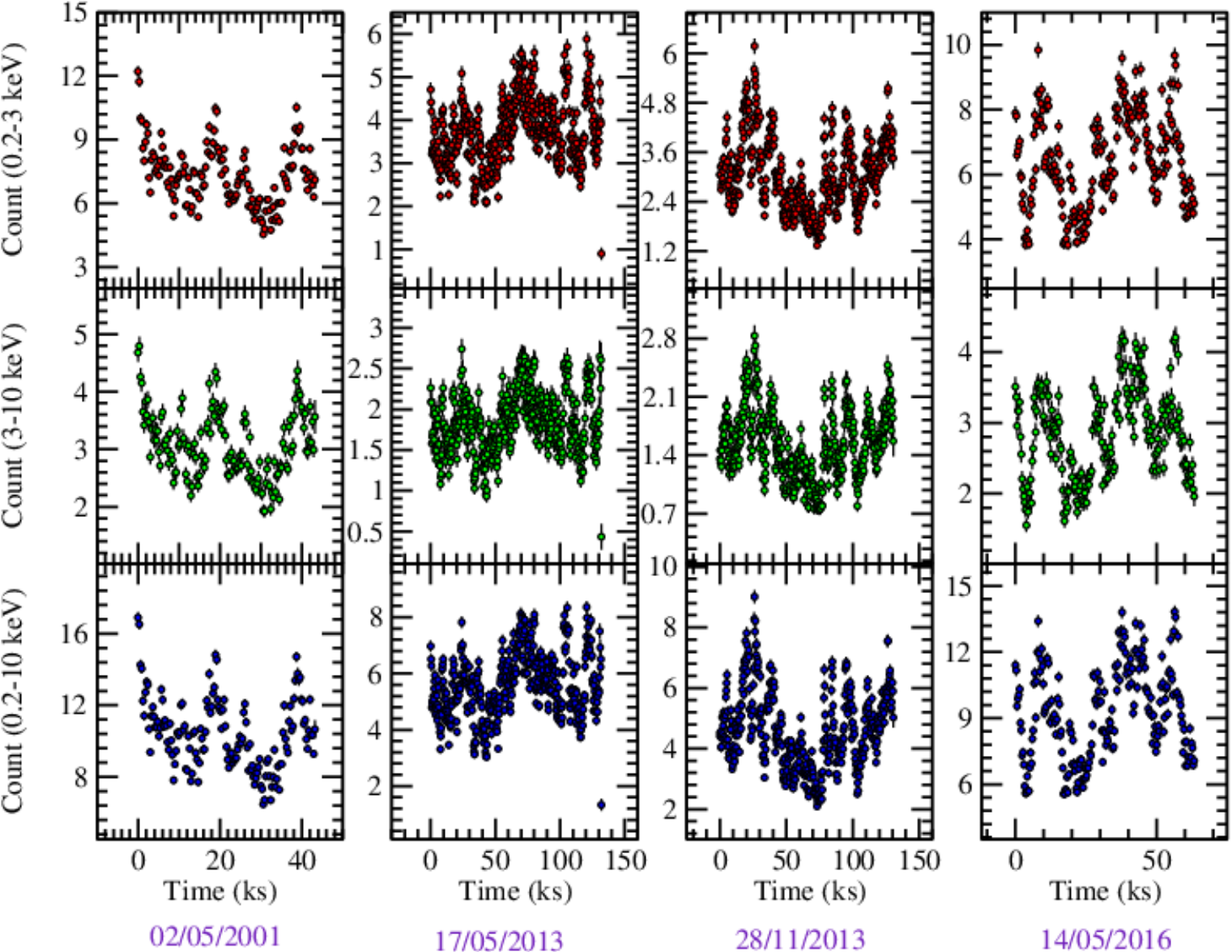}
    \caption{300 sec time binned light curves of four {\it XMM-Newton} data. The dates are mentioned at the bottom of each column.
    The three rows represent $0.2-3$ keV (soft; \textcolor{red}{red}), $3-10$ keV (hard; \textcolor{green}{green}), and $0.2-10$ keV (total; \textcolor{blue}{blue}) light curves respectively.}
    \label{fig:lcurv_xmm}
\end{figure*}

We produced $3-10$ keV, $10-78$ keV, and $3-78$ keV {\it NuSTAR} light curves to study the variability. The top, middle, and bottom panels of Figure~\ref{fig:lcurv} show the light curve in the (a) $3-10$ keV (soft band), (b) $10-78$ keV (hard band), and (c) $3-78$ keV (total) energy range, respectively. 
We have calculated the fractional variability ($F_{\rm var}$) of the three energy bands' light curves to study the variability \citep{Edelson2002,Vaughan2003}. The $F_{\rm var}$ of these soft, hard, and total bands light curves are found to be $0.270\pm0.003$, $0.196\pm0.006$, and $0.249\pm0.003$, respectively. 
We have also calculated the variability of the light curves of four {\it XMM-Newton} observations (see, Figure~\ref{fig:lcurv_xmm}). All the values of F$_{\rm var}$ for different energy ranges are given in Table~\ref{tab:table2}. We notice that the soft band (0.2-3 keV for {\it XMM-Newton} or 3-10 keV for {\it NuSTAR}) variability is always greater than the hard band (3-10 keV for {\it XMM-Newton} or 10-78 keV for {\it NuSTAR}) variability. Also, a consistent increase in the variability can be seen from 2001 to 2013 {\it XMM-Newton} observations, while it decreases after that in the 2016 observation. 

\begin{table*}
	\centering
	\caption{NGC 7314: Variability of light curves (binsize=300sec).}
	\label{tab:table2}
	\begin{tabular}{lccccccr} 
		\hline
		Satellite/ & Obs Id & Date & &&F$_{\rm var}$&&\\\cmidrule{4-8}  
		Instrument &        & (dd-mm-yyyy) & $0.2-3$ keV & $3-10$ keV & $0.2-10$ keV& $10-78$ keV& $3-78$ keV\\
		\hline
		{\it NuSTAR} & 60201031002 & 13-05-2016 & - &$0.270\pm0.003$ & - &$0.196\pm0.006$&$0.249\pm0.003$\\
        {\it XMM-Newton} & 0111790101 & 02-05-2001 &$0.191\pm0.002$ &$0.184\pm0.003$ &$0.188\pm0.002$ &-&-\\
                   & 0725200101 & 17-05-2013 &$0.202\pm0.002$ &$0.197\pm0.003$ &$0.200\pm0.002$ &-&-\\
                   & 0725200301 & 28-11-2013 &$0.282\pm0.002$ &$0.262\pm0.003$ &$0.274\pm0.002$ &-&-\\
                   & 0790650101 & 14-05-2016 &$0.223\pm0.002$ &$0.219\pm0.003$ &$0.222\pm0.002$ &-&-\\
		\hline
	\end{tabular}
\end{table*}

\begin{figure}
	\includegraphics[width=8.5cm]{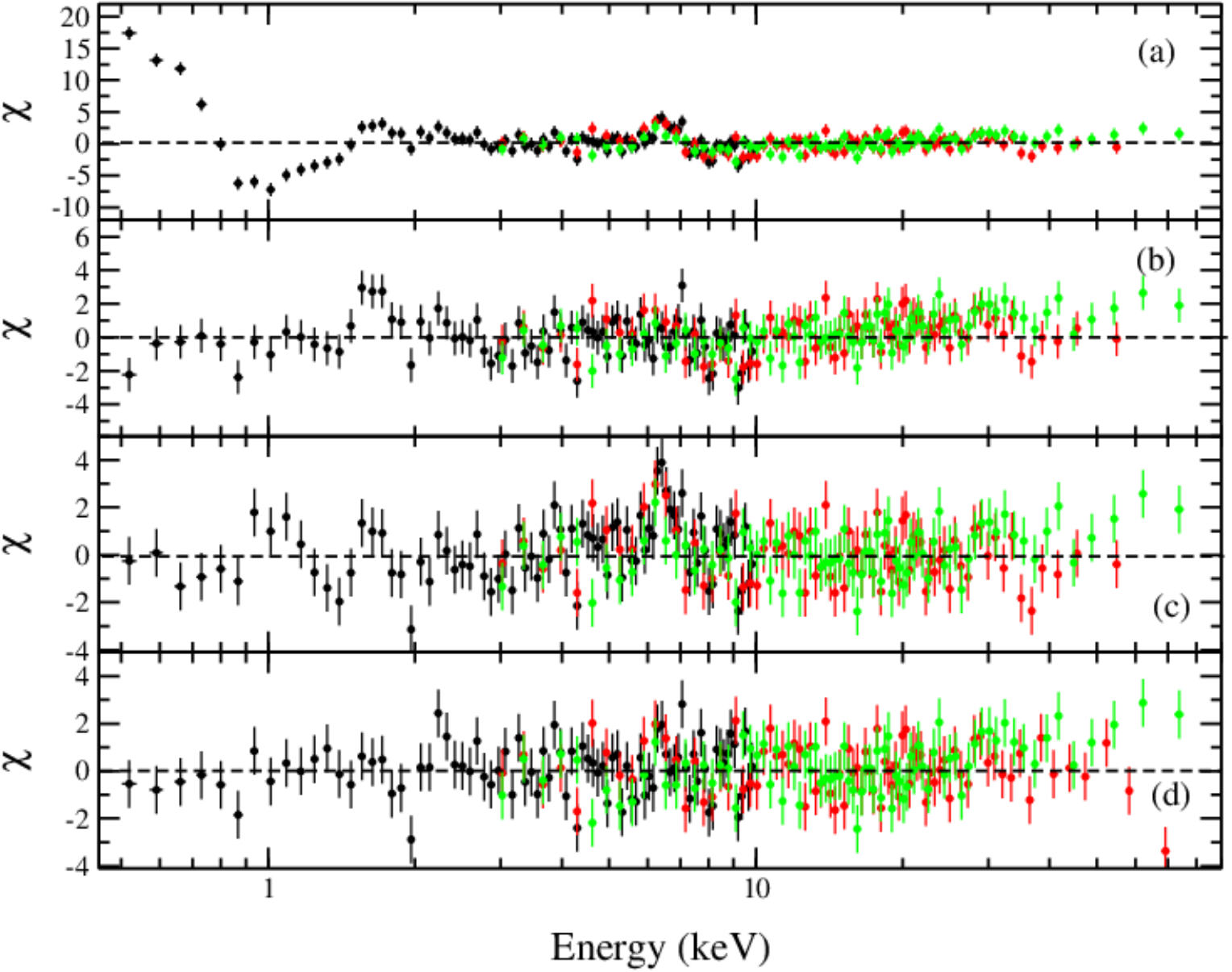}

    \caption{
    Variation of $\chi$ ((data-model)/error) for (a) Model~1 (\textsc{Tbabs*zTbabs*zCutoffPL}), (b) Model~2 (\textsc{Tbabs*zTbabs*(Bbody+zGaussian+zCutoffPL)}), (c) Model~3 (\textsc{Tbabs*zTbabs*(relxill+Bbody)}), and (d) Model~4 (\textsc{Tbabs*zTbabs*Gabs*(relxill+Bbody+xillver)}). The \textcolor{black}{black}, \textcolor{red}{red}, and \textcolor{green}{green} represent the $0.5-10$ keV {\it XMM-Newton}, $3-78$ keV {\it NuSTAR} FPMA and FPMB spectrum respectively.
    }
    \label{fig:chi}
\end{figure}

\begin{figure}
	\includegraphics[width=8.5cm]{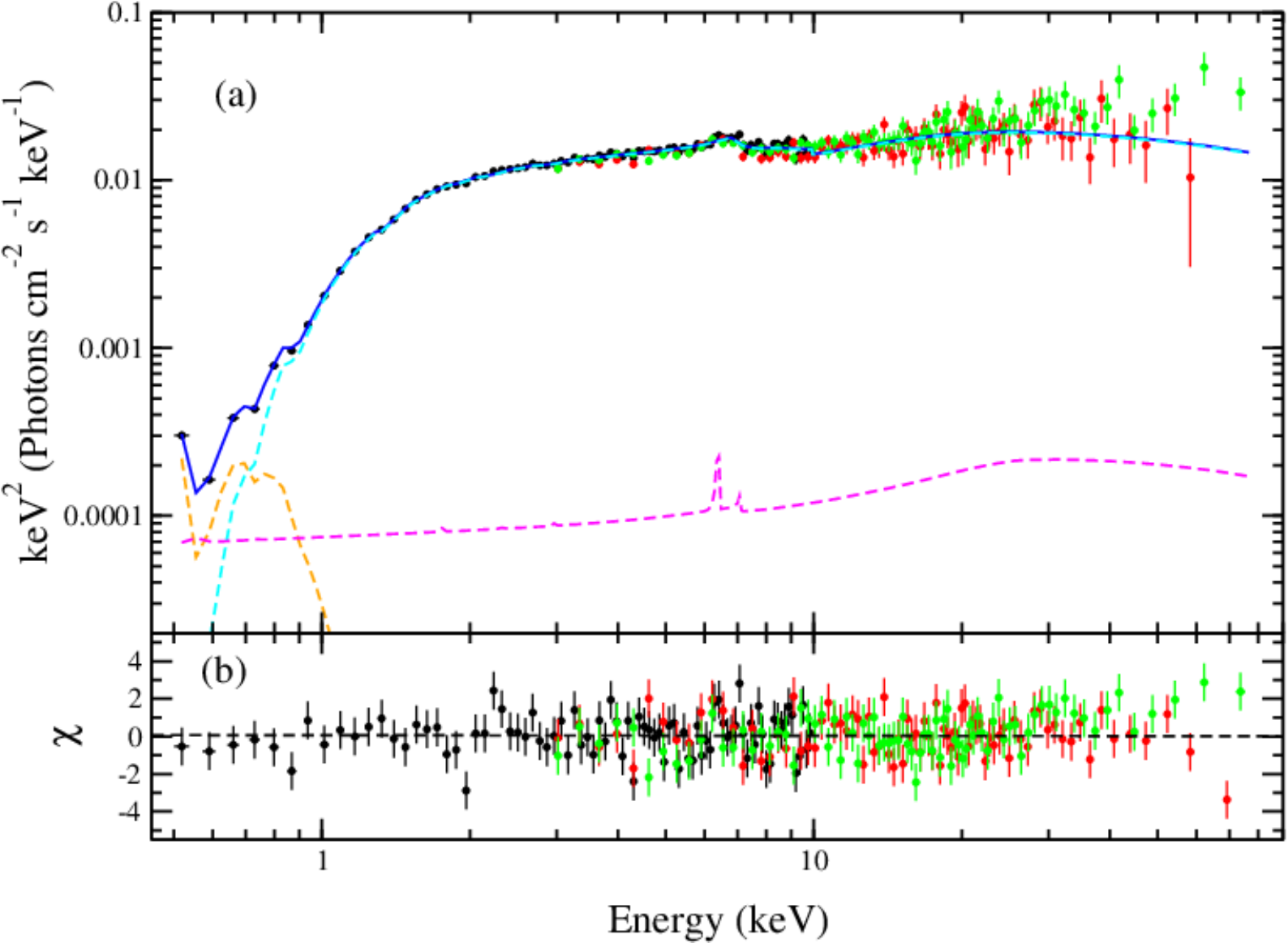}

    \caption{
    (a) Unfolded spectrum of \textsc{relxill+Bbody+xillver}. (b) $\chi^2$ variation of the upper panel spectrum. The \textcolor{black}{black}, \textcolor{red}{red}, and \textcolor{green}{green} represent the $0.5-10$ keV {\it XMM-Newton}, $3-78$ keV {\it NuSTAR} FPMA and FPMB spectrum respectively. The dashed \textcolor{orange}{orange},
    \textcolor{cyan}{cyan}, and \textcolor{magenta}{magenta} lines indicate the \textsc{Bbody}, {\tt relxill}, and \textsc{xillver} component respectively. The \textcolor{blue}{blue} line represents the combined model fitted spectrum.
    }
    \label{fig:spec}
\end{figure}

\subsection{Spectral analysis}
We use HeaSoft's spectral analysis package {\tt XSPEC}\footnote{https://heasarc.gsfc.nasa.gov/xanadu/xspec/}\citep{Arnaud1996} version 12.12.1 to fit the data. We make use of two \textsc{Tbabs} absorption models throughout our study for the line-of-sight absorption. Two absorption models are used to represent the Galactic and intrinsic line-of-sight absorption. The Galactic absorption was fixed at $1.45\times10^{20}$ cm$^{-2}$\citep{Dickey&Lockman1990}. We use \textsc{vern} scattering \citep{Verner1996} and \textsc{wilm} abundances \citep{Wilms2000}. The line of sight column density ($N_{\rm H}$) is kept free during the spectral fitting. 
$\chi^2$ statistic is used to determine the goodness of the fits. We use multiplicative model component \textsc{constant} as cross-normalization between the different spectra. We fixed the \textsc{constant} parameter at the unit value for the first spectrum while letting it vary for other spectra for simultaneous fit. \textsc{Tbabs} and \textsc{zTbabs} are used for the Galactic and intrinsic absorption.

\subsubsection{XMM-Newton and NuSTAR}
For the spectral analysis, we use combined {\it XMM-Newton} EPIC-pn and {\it NuSTAR} spectra in the $0.5-78$~keV energy range. Only the latest-epoch {\it XMM} observation is simultaneous with the only {\it NuSTAR} observation and thus used for the combined spectral analysis. For the spectral study, we use several phenomenological and physical models in this study.

\begin{itemize}

\item Model 1 :
We start our analysis with a simple model consisting of an absorbed power-law with a high energy cutoff. The model reads in \textsc{xspec} as \textsc{constant*Tbabs*zTbabs*zCutoffPL}. We obtain a photon index $\Gamma= 1.81 \pm 0.01$ with $\chi^2_{\rm red}$ 4582.44/1479. We notice a big bump below 1 keV and a signature of the Fe emission line around 6.5 keV in residual. The residual is shown in Figure~\ref{fig:chi}(a).

\item Model 2 :
We include \textsc{Bbody} model to incorporate the bump at soft energies. A \textsc{Gaussian} model is also added for the Fe K$\alpha$ line emission. The combined model is now:
\textsc{constant*Tbabs*zTbabs*(Bbody+zGaussian+zCutoffPL)}. We obtain a photon index ($\Gamma$) of $1.85\pm0.01$. The \textsc{Bbody} model gives a temperature of $k\rm T$ $0.05\pm0.001$ keV. The \textsc{Gaussian} fit gives line energy of $6.51\pm0.04$ keV and width ($\sigma$) of $0.31\pm0.04$ keV. We obtain $\chi^2$/dof = 2027/1474 from the best-fit spectrum. A prominent signature of reflection can be noted in the residual (see, Fig.~\ref{fig:chi}(b)). We also see significant variation in the lower energy spectrum. The model parameters are given in Table~\ref{tab:table3}.

\item Model 3 :
As the reflection hump is seen in the residual of Model-2 fit, next, we employed a relativistic reflection model for the spectral analysis. Given that the X-ray spectrum of active galactic nuclei (AGN) comprises direct and reflected emissions originating from the accretion disk and the irradiation of a fraction of primary X-rays on the disk, the extent of reflection can be deduced through the ratio of direct and reflected flux. This concept is encapsulated within the relativistic reflection model \textsc{relxill} \citep{Garcia2014}, which integrates the \textsc{xillver} reflection code with the relativistic line profiles code \textsc{relline} \citep{Dauser2016}. In this model, the reflection fraction (RF) denotes the ratio of photons impacting the disc to those reaching infinity. The accretion disc spans from the marginally stable radius ($R_{\rm in} = 1.24~r_g$) to $R_{\rm out} = 1000~r_g$, and relativistic light bending phenomena can give rise to a warped disk appearance. \textsc{relxill}, functioning as the standard relativistic reflection model, characterizes accretion irradiation via a broken power-law emissivity. The ionization states of the accretion disc encompass a range from log~$\xi$ = 0 (neutral) to log~$\xi$ = 4.7 (highly ionized), while the iron abundance ($A_{\rm Fe}$) of disk material is expressed in solar abundance units.
We fit the combined spectra with the \textsc{relxill} model (\url{http://www.sternwarte.uni-erlangen.de/~dauser/research/relxill/}) to incorporate the reprocessed emission. The spectral fit with \textsc{relxill} gives us an acceptable fit. However, still, a positive residual is seen at the soft energy band ($>1$~keV). Hence, we add a \textsc{Bbody} for the soft energy bump. The combined model is: \textsc{constant*Tbabs*zTbabs*(relxill+Bbody)}. During the fitting, we fixed the outer disk radius at 1000 $\rm R_g$. The iron abundance is fixed at the solar value. The \textsc{Bbody} temperature fits at $k\rm T$ $\sim$ $0.05$ keV. The emissivity indices ($q_1$ and $q_2$) for the coronal flavor models (as $r^{-q_1}$ between R$_{\rm in}$ and $R_{\rm br}$) gives maximum value of $q_1\sim$10 while $q_2$ (as $r^{-q_2}$ between $R_{\rm br}$ and $R_{\rm out}$) is fixed at 3. The inner disk radius is obtained to be $1.30^{+0.28}_{-0.23}$ $R_{\rm ISCO}$ (inner stable circular orbit). The inner disc inclination angle is found to be $\sim44^\circ\pm2.19$. The photon index ($\Gamma$) is $1.86\pm0.01$. The ionization parameter from the \textsc{relxill} fit at log$\xi$ $\sim3.14\pm0.03$. We notice that a small signature of emission is still present around $\sim 6-7$ keV. Also, the residual shows a dip around 1-2 keV. The variation of $\chi$ is shown in Fig.~\ref{fig:chi}(c).

\item Model 4 :
We include an absorption multiplicative component \textsc{Gabs} to incorporate the dip around $\sim 1-2$ keV. Also, we add a \textsc{xillver} component for the emission around $\sim 6-7$ keV for the narrow line emission. These two inclusions improve the fit statistics Fig.~\ref{fig:chi}(d). The combined model is: \textsc{constant*Tbabs*zTbabs*Gabs* (relxill+Bbody+xillver)}. The parameters of \textsc{xillver} model are fixed with \textsc{relxill} model parameters except for the model normalization. We have also fixed the value of the logarithmic ionization parameter (log$\xi$) at 0 to consider the contribution of the neutral iron emission line. The spectral analysis with this model gave us a good fit with $\chi^2$/dof= 1713/1467. The \textsc{gabs} component gives line energy, width, and strength of the absorption dip to be $1.35\pm0.03$ keV, $0.12\pm0.01$ keV, and $0.02\pm4e-3$ respectively.  The \textsc{relxill} model fit parameters found as: photon index($\Gamma$), inclination angle, spin, inner disk radius ($R_{\rm in}$) are $1.88\pm0.01$, $\sim44^\circ~^{+1.24}_{-1.32}$, $0.71^{+0.14}_{-0.21}$, $1.34^{+0.25}_{-0.33}$ $R_{\rm ISCO}$ respectively. The temperature obtained from \textsc{Bbody} model is $0.05\pm0.002$ keV. The best-fit unfolded-spectrum with the $\chi$ variation is shown in Fig.~\ref{fig:spec}(a-b). The \textsc{Bbody}, \textsc{relxill}, and \textsc{xillver} models are represented by orange, cyan, and magenta colors, respectively. The blue line indicates the combined model curve. The {\it XMM-Newton}, {\it NuSTAR}/FPMA, and {\it NuSTAR}/FPMB data are portrayed by black, red, and green color points, respectively. The parameters of the fitted model are given in Table~\ref{tab:table3}.

\end{itemize}

\begin{table}
	\centering
	\caption{Spectral Results for combined {\it XMM-Newton} and {\it NuSTAR} data.}
	\label{tab:table3}
	\begin{tabular}{lccr} 
		\toprule
		Model & Parameters & Values & $\chi^2$/dof\\
		\hline
	  Model 2&&&\\\cmidrule{1-1}
	    \textsc{CutoffPL} & $\Gamma$ & $1.85^{+0.01}_{-0.01}$ & 2027/1474\\
		 & $E_{\rm cut}$ (keV) & $483\pm79$ &\\
		 & $Norm$ ($\times10^{-4}$)& $120\pm1$ & \\
		 \textsc{Gaussian}& $E_{\rm line}$ (keV) & $6.51^{+0.04}_{-0.04}$ & \\
		 & $\sigma$ (KeV)   & $0.31^{+0.04}_{-0.04}$ &\\
		 & $Norm$  ($\times10^{-6}$)   & 40$\pm$3 &\\
      \textsc{Bbody}& $k$T (keV) & 0.05$\pm$0.001 &\\
      & $Norm$ & $0.05^{+0.01}_{-0.01}$ &\\
      \textsc{Tbabs}& $N_{\rm H}$ ($\times10^{22}~{\rm cm^{-2}}$) & $1.05^{+0.01}_{-0.01}$&\\
      \textsc{Constant}& $C_{\rm pn/FPMA}$ & $0.93\pm0.01$ &\\
                       & $C_{\rm pn/FPMB}$ & $0.97\pm0.01$ &\\
\hline
    Model 4&&&\\\cmidrule{1-1}
      \textsc{relxill}& $q_1$ & $10^a$ & 1713/1467\\
      & $q_2$ & $3^a$ &\\
      & $R_{\rm br}$ & $12^a$ &\\
      & $a$ & $0.71^{+0.14}_{-0.21}$ &\\
      & inclination ($\theta^\circ$)& $43.91^{+1.34}_{-1.32}$ &\\
      & $R_{\rm in}$ (ISCO) & $1.34^{+0.25}_{-0.33}$ &\\
      & $R_{\rm out}$ (r$_{\rm g}$) & $1000^a$ &\\
      & $\Gamma$ & $1.88^{+0.01}_{-0.01}$ & \\
      & $Fe_{\rm abund}$ & $1^a$ &\\
      & $E_{\rm cut}$ & $300^a$&\\
      & ${\rm log}~\xi$ & $3.12^{+0.03}_{-0.03}$ &\\
      & $refl_{\rm frac}$ & $1.32^{+0.11}_{-0.11}$ &\\
      & $Norm$ ($\times10^{-6}$) & 151$\pm$8  &\\
      \textsc{Bbody}& $kT$ (keV) & 0.05$\pm$0.002 &\\
      & $Norm$ & $0.06^{+0.01}_{-0.01}$ &\\
      \textsc{gabs}& $E_{\rm abs}$ & $1.35^{+0.03}_{-0.03}$ & \\
      & $\sigma$ & $0.12^{+0.01}_{-0.01}$ &\\
      & $Strength$ & 0.02$\pm$0.004 &\\
     \textsc{xillver} & ${\rm log}~\xi$ & $0^a$ &\\
      & $Norm$ ($\times10^{-6}$)    & 1$\pm$0.8 &\\
      \textsc{Tbabs}& $N_{\rm H}$ ($\times10^{22}~{\rm cm^{-2}}$) & $1.17^{+0.02}_{-0.01}$ &\\
      \textsc{Constant}& $C_{\rm pn/FPMA}$ & $0.93\pm0.01$ &\\
                       & $C_{\rm pn/FPMB}$ &  $0.98\pm0.01$ &\\
      \bottomrule
	\end{tabular}
 \noindent{Best-fit parameters of Model 2 and Model 4 are given. The superscript $a$}
 \noindent{indicates that the parameters are kept fixed at their given value during the fit.}
 \noindent{The $\pm$ values represent the errors with 90\% confidence. The errors for some} 
 \noindent{parameters of Model 4 estimated using the MCMC method are given as a} 
 \noindent{corner plot in Fig.\ref{fig:corner} (APPENDIX).}
\end{table}

\subsubsection{XMM-Newton}

Four {\it XMM-Newton} observations (ObsId1: 0111790101 on 02/05/2001, ObsId2: 0725200101 on 17/05/2013, ObsId3: 0725200301 on 28/11/2013, ObsId4: 0790650101 on 14/05/2016) is taken for the spectral study. We fit the four EPIC-pn spectra of {\it XMM-Newton} simultaneously with \textsc{Tbabs*zTbabs*Gabs(relxill+Bbody+xillver)}. We keep the inclination angle and spin parameter ($a$) linked for all four spectra. Also, we have pegged the parameters of the \textsc{xillver} model to follow \textsc{relxill} model parameters except for the normalization. The iron abundance is kept fixed at solar value. Other parameters are kept free to vary for individual spectra. The high energy cutoff for \textsc{relxill} model is fixed at 100~keV. From the best-fitted results, we have obtained a common spin parameter ($a=0.65^{+0.12}_{-0.13}$) and inclination angle ($\theta=44.72^\circ~^{+2.15}_{-1.81}$) from all these different epochs spectra. We have noticed satistically marginal variation of the centroid energy and width for the four epochs.
The values (parameters) of $r_{\rm ISCO}$, $\Gamma$, and \textsc{Gabs} model have been given in the first part of Table~\ref{tab:table4}.

To estimate the variation of the contribution of both broad and narrow iron line emission, we fit the four {\tt XMM-Newton} spectra with \textsc{Tbabs*zTbabs*Gabs(zCutoffPL+Bbody+zGa+zGa)} model. First, we include only one \textsc{Gaussian} with all the parameter values kept free during the fitting. The obtained results are given in the second part of Table~\ref{tab:table4}. We notice a small signature of emission line $\sim$6.4 keV after the fit. So, we add one more \textsc{Gaussian} with a fixed line width of 0.01 keV to incorporate this line. The centroid width and normalization of the narrow component are given in Table~\ref{tab:table4}. During this spectral fitting with phenomenological models, we keep the \textsc{Gabs}, \textsc{Bbody} model parameters in a small range averaged around the values obtained from the physical model's fitted values.
\begin{table*}
	\centering
	\caption{Spectral Results for simultaneously fitted four {\it XMM-Newton} EPIC-pn spectra.}
	\label{tab:table4}
	\begin{tabular}{lcccr} 
		\hline
  parameters & ObsId. 1 & ObsId. 2 & ObsId. 3 & ObsId. 4\\
  & 02/05/2001 & 17/05/2013 & 28/11/2013 & 14/05/2016\\
  \hline
\multicolumn{3}{l}{\textsc{Tbabs*zTbabs*Gabs(relxill+xillver+Bbody)}}\\\cmidrule{1-3}

       $R_{\rm in}$ (ISCO) & $1.52^{+0.08}_{-0.09}$ & $2.18^{+0.23}_{-0.39}$ & $1.79^{+0.11}_{-0.09}$&$1.29^{+0.09}_{-0.10}$\\
       $\Gamma$ & $1.93^{+0.02}_{-0.01}$ & $1.87^{+0.01}_{-0.01}$& $1.76^{+0.02}_{-0.03}$ & $1.84^{+0.02}_{-0.02}$ \\
       $E_{\rm abs}$ (keV) & $1.37^{+0.05}_{-0.04}$ & $1.32^{+0.04}_{-0.04}$& $1.32^{+0.03}_{-0.03}$& $1.35^{+0.05}_{-0.05}$ \\
       line width (keV) & $0.09^{+0.04}_{-0.05}$ & $0.11^{+0.03}_{-0.03}$& $0.10^{+0.06}_{-0.04}$& $0.08^{+0.05}_{-0.04}$ \\
       $Strength$ & 0.01$\pm$0.003 & 0.01$\pm$0.002& 0.02$\pm$0.004& 0.01$\pm$0.003 \\
       $N_{\rm H}$ ($\times10^{22}~{\rm cm^{-2}}$) & $1.13^{+0.02}_{-0.02}$ &$1.19^{+0.01}_{-0.01}$ &$1.10^{+0.02}_{-0.01}$ &$1.14^{+0.03}_{-0.03}$\\
       \hline
\multicolumn{3}{l}{\textsc{Tbabs*zTbabs*Gabs(zCutoffPL+Bbody+zGa+zGa)}}\\\cmidrule{1-3}
     $E_{\rm gaussian}^{\rm broad}$ (keV) & 6.54$\pm$0.07 & 6.54$\pm$0.02 &6.46$\pm$0.04 &6.62$\pm$0.06 \\
		line width, $\sigma$ (keV)      & 0.40$\pm$0.07 &0.33$\pm$0.05 &0.24$\pm$0.07 &0.37$\pm$0.06 \\  
       EW (keV)      & $\sim$0.20 &$\sim$0.15 &$\sim$0.14 &$\sim$0.19 \\
       $\nu_{\textsc{fwhm}}^{\rm broad}$ (km~$\rm s^{-1}$)      & 18,300$\pm$3300 &15,300$\pm$2400 &11,400$\pm$3300 &16,800$\pm$2700 \\
		 $Norm$  ($\times10^{-6}$)  &80$\pm$8 & 40$\pm$4 &30$\pm$3&70$\pm$6\\
   $R_{\rm Fe~K\alpha}$ ($10^{14}$~cm)    & $\sim$2.57 & $\sim$3.67 & $\sim$6.62 & $\sim$3.05 \\
     $E_{\rm gaussian}^{\rm narrow}$ (keV) & 6.38$\pm$0.06 & 6.37$\pm$0.02 &6.46$\pm$0.02 &6.38$\pm$0.01 \\
       $\nu_{\textsc{fwhm}}^{\rm narrow}$ (km~$\rm s^{-1}$)      & 47$\pm$0.45 &47$\pm$0.15 &47$\pm$0.15 &47$\pm$0.09 \\
   $Norm$  ($\times10^{-6}$)   & 6$\pm$1 &10$\pm$2&10$\pm$2&10$\pm$3\\       
   \hline
   $L_{2-10~{\rm keV}}$ ($\times10^{42}~{\rm erg~s}^{-1}$) & 1.96$\pm$0.02 & 1.17$\pm$0.02 & 0.97$\pm$0.03 & 1.79$\pm$0.05\\
   \hline
      \end{tabular}
      
     \noindent{The errors are calculated using {\tt fit err} command and represent 90\% confidence level. The Observation dates are written in dd/mm/yyyy format.}
\end{table*}

\subsubsection{RXTE/PCA data}
\begin{figure}
	\includegraphics[width=8.5cm]{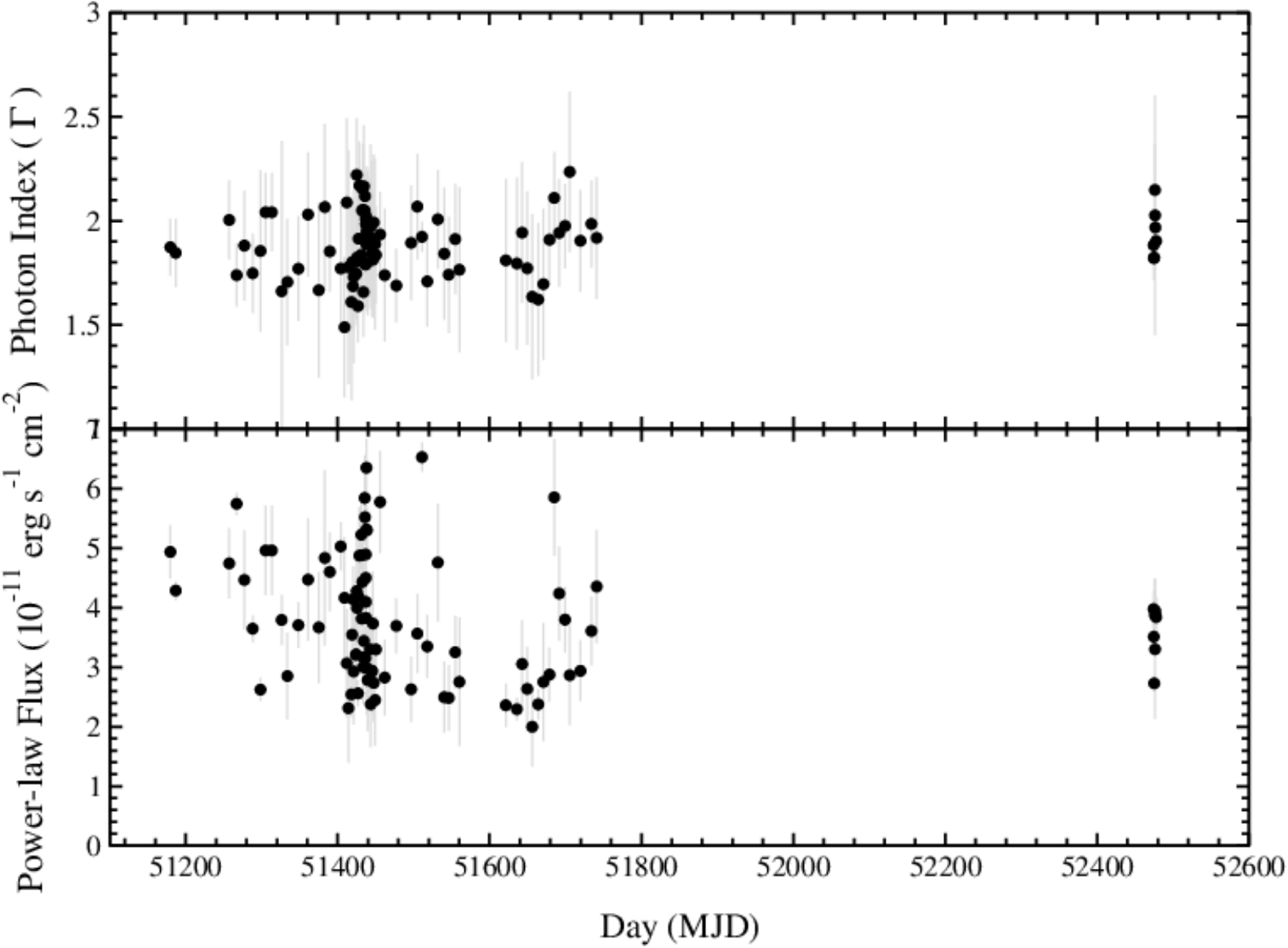}
    \caption{Variation of photon indices ($\Gamma$) and power-law flux (in $10^{-11}$ $\rm erg~s^{-1}~cm^{-2}$) with MJD (for {\it RXTE}/PCA data).
    }
    \label{fig:rxte_pl}
\end{figure}

\begin{figure}
	\includegraphics[width=8.5cm]{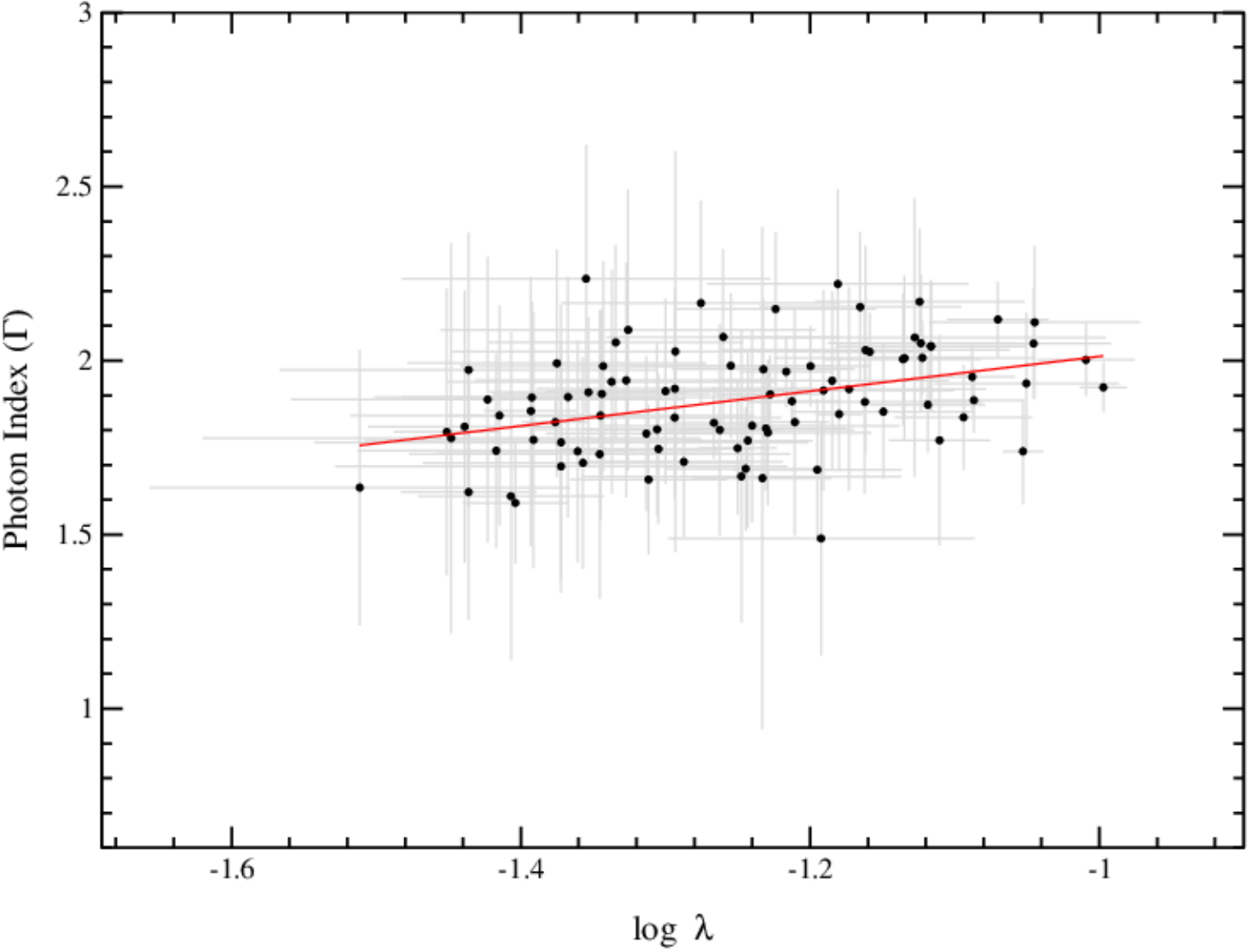}
    \caption{Variation of photon indices ($\Gamma$) with log~$\lambda$ for {\it RXTE}/PCA data. $\lambda$ is the ratio between L$_{\rm bol}$ and L$_{\rm Edd}$. L$_{\rm bol}$ is considered to be 20$\times$L$_{2-10~{\rm keV}}$ \citep{Vasudevan&Fabian2009,Duras2020}.
    }
    \label{fig:corln}
\end{figure}

We analyze total 85 {\it RXTE}/PCA observations of NGC 7314 from January 01 1999 (MJD=51179.74) to July 16 2000 (MJD=51741.09) and 7 {\it RXTE}/PCA observations from July 19, 2002 (MJD=52474.22) to July 22, 2002 (MJD=52477.71) for studying the variation for a longer period. We use the combined \textsc{tbabs*ztbabs*powerlaw} model to fit the $3-20$ keV spectra during this period. Whenever a significant contribution of iron emission line is noticed, we included \textsc{Gaussian} model at around 6.5 keV to incorporate the Fe K$\alpha$ emission. The total studied observations are 92 (85+7). The photon index ($\Gamma$) varies from 1.48$\pm$0.34 to 2.23$\pm$0.38 during our observation period. We calculate $2-10$ keV unabsorbed power-law flux using {\tt cflux} command in {\tt XSPEC}. We notice a variation of flux from 1.99$\pm$0.67 to 6.53$\pm$0.24 $\times10^{11}~erg~s^{-1}~cm^{-2}$. We examined the factuality of photon index and flux change by checking whether they are consistent with a constant in time.
The best constant fit to the photon index yields a $\chi^2_\nu$ of 0.45, while that to the flux has a $\chi^2_\nu$ of 7.76.
Although the $\Gamma$s do not show obvious variation, given its
current uncertainties, the flux variation is seen at a high
significance. The variation of photon index ($\Gamma$) and power-law flux (in $2-10$ keV range) are given in Figure~\ref{fig:rxte_pl}.
On the other hand, a weak correlation is found between the $\Gamma$
and power-law flux, with Pearson’s correlation coefficient of 0.38, corresponding to a p-value of 0.0002. It hints that the $\Gamma$ indeed varies. We obtain the rank coefficient from the Spearman rank correlation as 0.42 with a p-value of $4\times10^{-5}$ (0.00004). This suggests a moderate positive correlation, and the result is statistically significant. We also calculated the Eddington ratio ($\lambda=\frac{{L_{\rm bol}}}{{L_{\rm Edd}}}$) and performed a linear fit of $\log \lambda$ vs. $\Gamma$ as shown in Fig.~\ref{fig:corln}.
$L_{\rm bol}$ is considered to be $20\times L_{\rm 2-10~keV}$\citep{Vasudevan&Fabian2009,Duras2020}. 
 This relationship between $\Gamma$ and the Eddington ratio has been a subject of interest in understanding the accretion processes in AGN.

\section{Discussions}

\subsection{Variability}
We study the accretion properties of narrow-line Seyfert NGC 7314 using {\it RXTE}, {\it XMM-Newton}, and {\it NuSTAR} data. The timing analysis conducted on NGC 7314 using {\it NuSTAR} revealed significant variability across different energy bands ($F_{\rm var}$ $\sim$ $0.270\pm0.003$ (3-10 keV), $0.196\pm0.006$ (10-78 keV), and $0.249\pm0.003$ (3-78 keV)). We also estimate the variability of light curves from {\it XMM-Newton} data. For both the case of {\it NuSTAR} and {\it XMM-Newton}, we see a similar pattern on the light curve in the soft and hard bands. 
We notice that both in the soft (0.2-3 keV) and hard (3-10 keV) energy bands in the {\it XMM-Newton} light curves, the variability increases (0.191$\pm$0.002 to 0.282$\pm$0.002 in soft; 0.184$\pm$0.003 to 0.262$\pm$0.003 in hard) from 2001 to 2013. In 2016, the variability again showed lower values both in the soft ($\sim$0.223$\pm$0.002) and hard energy bands ($\sim$0.219$\pm$0.003). Also, we notice that the variability in the soft energy bands (i.e., 0.2-3 keV for {\it XMM-Newton} and 3-10 keV for {\it NuSTAR}) is always higher than the variability in the hard energy bands (i.e., 3-10 keV for {\it XMM-Newton} and 10-78 keV for {\it NuSTAR}), which is more significantly verified in case of the {\it NuSTAR} observation. This implies that the high variability mainly comes from the soft X-ray emitting region. The fractional variability ($F_{\rm var}$) values obtained for the soft, hard, and total energy bands were consistent with previous studies and indicated the presence of intrinsic variability in the source. We also notice similar types of variation in both the soft and hard energy bands. These results suggest the existence of dynamic processes within NGC 7314, such as accretion disk instabilities or changes in the coronal emission region. A possible explanation could be that the primary soft X-ray continuum is a variable source, produced in some hot corona closer to the SMBH, while the high energy photons are produced from the scattering of these primary X-ray photons by a constant temperature high energy cloud, located away from the SMBH (see also, \cite{Lawrence1985} for spectral changes in NGC 4051). The variable soft-X rays thus get into multiple scattering and come out as more smooth-amplitude high-energy light curves. Also, the increase and decrease in the values of the $F_{\it var}$ indicates a change in the inner accretion properties that happened during the 2001 to 2013 period, and the reverse phenomenon happened during the 2013 to 2016 period.

\subsection{Spectral Evolution}
We analyze combined {\it XMM-Newton} and {\it NuSTAR} data for a broadband study of the accretion properties of NGC 7314. In the spectral analysis, we employed various models (a combination of phenomenological and physical) to unravel the emission components and their evolution over time. 

The combined spectrum of {\it XMM-Newton} and {\it NuSTAR} is best fitted with model 4. From the combined spectral fit, we obtain a low inner edge of the accretion disk, a moderate inclination angle, and a relatively high spin value. The spectrum shows a clear signature of an absorption component around $1.35$~keV and a soft excess component with a peak around 0.05~keV. Fe~K$\alpha$ lines come from two different regions- one from some high ionization region and one from a low (or neutral) ionization region. 

To study the evolution, we study the four XMM-Newton spectra (see, Table\ref{tab:table4}). We obtain spin parameter ($a\sim0.65^{+0.12}_{-0.13}$) and inclination angle ($\theta\sim45^\circ~^{+2.15}_{-1.81}$) from the simultaneous fitting of the XMM observations. The spectral analysis of four {\it XMM-Newton} data revealed a change in the accretion properties. These variations could be attributed to changes in the accretion flow, the geometry of the emitting regions, or variations in the intrinsic source properties. In Table~\ref{tab:table4}, although not significant for all, a few parameters show similar values in the 2001 and 2016 observations. These changes indicate a mild state transition from 2001 to 2013 and retracing back again to 2016.

The change in the inner disk radius ($R_{\rm in}$) and absorption component \textsc{Gabs} is noticed during this time. Although the photon index ($\Gamma$) has not shown significant change, a rough pattern can be noticed similar to the other parameters. The 2-10 keV luminosities ($L_{2-10 keV}$) also show a similar feature. 

The variation of the Fe~K$\alpha$ line emission and the absorption feature is also noticed in the four {\it XMM-Newton} data (May 2, 2001; May 17, 2013; November 28, 2013, and May 5, 2016). 
The equivalent width of the broad Fe~K$\alpha$ line also shows variation during this period. From 2001 to 2013, the equivalent width increased (from 47 eV to 56/66 eV ), then in 2016, it showed a lesser value ($\sim$36 eV).

The long-term analysis of {\it RXTE}/PCA data provides insights into the temporal variations of NGC 7314. The weak positive correlation suggests that the $\Gamma$ also increases slightly as the power-law flux increases. This behavior implies that the spectrum becomes softer with increasing flux. This result aligns with the general trend observed in other AGNs, where a positive correlation between $\Gamma$ and the Eddington ratio ($\lambda$) is often found. Previous studies have shown varying degrees of correlation between $\Gamma$ and the $\lambda$. For example, \citet{Shemmer2006} found a strong positive correlation between $\Gamma$ and $\lambda$ in a sample of AGNs, suggesting that higher accretion rates (higher $\lambda$) are associated with softer spectra (higher $\Gamma$). \citet{Brightman2013} also reported a positive correlation between $\Gamma$ and the $\lambda$ in a sample of AGNs, reinforcing the idea that the accretion rate influences the spectral shape. However, \citet{Risaliti2009} noted that the correlation can vary significantly depending on the sample and the specific characteristics of the AGNs studied, indicating that other factors, such as the black hole spin and the geometry of the accretion disk, can also play a significant role.

Our findings for NGC 7314, although showing a weaker correlation compared to some studies, are still consistent with the general trend observed in AGNs. The weak correlation might be due to intrinsic variability in NGC 7314 or differences in the physical conditions of the accretion flow compared to other AGNs. In summary, the positive correlation between $\Gamma$ and the Eddington ratio ($\lambda$) in NGC 7314 supports the idea that higher accretion rates lead to softer X-ray spectra. This is in line with previous studies, although the strength of the correlation in NGC 7314 appears to be weaker.

The pattern of this spectral evolution indicates that NGC 7314 could be a potential changing-state AGN. Although the variation of the spectral properties is not very significant for NGC 7314, a clear trend can be noticed during $\sim$15 years of observations. Almost similar properties can be noticed in Mrk 110 in 2001 when it completely did not shift to type-2, but in some intermediate Seyfert type, classified it as moderately changing state AGN \citep{Porquet2024}.

\subsection{Absorption}{\label{absorption}}
An absorption component \textsc{Gabs} is required to fit the spectra. The line energy of this absorption component is centered around $1.35\pm0.03$ keV with a line width of $0.12\pm0.01$ keV (see model 4). 

We also estimate the dust sublimation radius (inner radius of the dusty torus, $R_{\rm dust}$), following the methods of \cite{Nenkova2008a,Nenkova2008b},
\begin{equation}
    R_{\rm dust} = 0.4\left(\frac{L_{\rm bol}}{10^{45}{\rm erg~s}^{-1}}\right)^{1/2}\left(\frac{1500~{\rm K}}{T_{\rm sub}}\right)^{2.6}~{\rm pc},
\end{equation}
where $L_{\rm bol}$ and $T_{\rm sub}$ are the bolometric luminosity and dust sublimation temperature. $T_{\rm sub}$ is generally assumed to be the sublimation temperature of graphite grains, $T\sim1500~{\rm K}$ \citep{Kishimoto2007}. We consider the bolometric luminosity to be $L_{\rm bol}\sim20\times L_{2-10~{\rm keV}}$\citep{Vasudevan&Fabian2009,Duras2020}. We obtain the average dust sublimation radius from the four observations (see, last row of Table~\ref{tab:table4} for $L_{2-10}$ luminosity) to be 0.05~pc (or $\sim2\times10^{17}$~cm).
From a study of {\it XMM-Newton}, {\it Suzaku}, and {\it ASCA} data, a coherent depiction of the system's geometry within the framework of a unified model is given \citep{Ebrero2011}. The diverse observed properties are explained as neutral gas clouds moving across our line of sight. These clouds could possibly be responsible for the absorption feature around 1.3~keV. These clouds could possibly be located between the BLR and the dusty torus since a small variability can be observed in the absorption parameters during our studied period. However, from the line width of the absorption, it is evident that the absorption occurred in some high ionization regions and might be closer than the BLR. 

We also estimate the BLR radius ($R_{\rm BLR}$) from the X-ray luminosity \citep{Kaspi2005}.
\begin{equation}
    R_{\rm BLR} = 7.2\times10^{-3}\left(\frac{L_{2-10}}{10^{43}~{\rm erg~s}^{-1}}\right)^{0.532}~{\rm pc},
\end{equation}

where $L_{2-10}$ is the $2-10$~keV X-ray luminosity. We obtain the $R_{\rm BLR}$ from the four observations to be 9.08$\times10^{15}$~cm, 6.90$\times10^{15}$~cm, 6.24$\times10^{15}$~cm, 8.65$\times10^{15}$~cm respectively.

In a study \citep{Armijos-Abendano2022} with {\it XMM-Newton} observations of NGC 7314 (and several other sources) using the hardness-ratio curves, the time intervals in which the clouds are eclipsing the central X-ray source has been investigated. They estimated that the eclipsing clouds with distances from the X-ray emitting region of 9.6$\times~10^{15}~{\rm cm}$ (or 3.6$\times~10^{4}~{\rm r_g}$ considering $M_{\rm BH}=10^6$~$M_\odot$) are moving at Keplerian velocities $\sim$1122 km/s. The distance of the clouds ($\sim10^{16}$~cm) is similar to our estimated BLR radius. This indicates that the obscuring clouds are associated with BLR.
To justify the line-width absorption, a high ionizing region, even closer to the BLR being the origin of the absorption, can not be overruled.

\subsection{Soft excess}
The soft excess in the spectra is modeled with a \textsc{Bbody} component. The temperature ($kT\sim0.05$ keV) almost remained invariant for every combination of models. Even while fitting the different epoch {\it XMM-Newton} spectra, the temperature remains the same, except the normalization value was $\sim0.08\pm0.02$ for 2001 and 2016, and $\sim0.06\pm0.01$ for two 2013 observations. Although it is not certain about the origin of the soft excess \citep{Pravdo1981,Arnaud1985,Turner1989}, however, this excess emission below 2 keV in X-ray is very common in narrow-line Seyfert 1 AGNs. Classically, this soft excess is modeled with a black body emission with a temperature of 0.1-0.2 keV \citep{Walter&Fink1993,Czerny2003,Crummy2006}. However, the temperature of the soft excess is too high to be directly emitted from the standard accretion disk \citep{Shakura1973}. A narrow temperature range was obtained for a huge range of supermassive black holes (M$\sim10^6-10^8~M_\odot$) when a sample of AGNs was modeled with a Compton scattered disk component \citep{Gierlinski2004}. 
The observed soft excess in the spectra could possibly be coming from the multiple scattering of the relatively high-energy photons. The soft excess could arise from the hot corona when seed photons suffer less scattering \citep{Nandi2023}. The hot plasma, responsible for the low variability of the high-energy photons, could be the origin of the soft excess. 

\subsection{Iron emission line}
The production of the narrow Fe~K$\alpha$ line is often attributed to the reprocessing of central X-ray coronal emission by remote materials like the dusty torus \citep{Krolik1987,Nandra2006}. Conversely, the broad Fe~K$\alpha$ line is commonly thought to originate from the innermost section of the super-massive BH, most probably from the accretion disk or BLR \citep{Nandra1997,Zoghbi2014,Kara2015}. Its asymmetric profile is linked to relativistic beaming and gravitational redshift effects \citep{Fabian1989,Fabian2000}. Given the substantial distance and scale of the reprocessing material, it's plausible that the narrow component exhibits considerably less variability than the broader component. 

To estimate the parameters of these broad and narrow iron line features, we fit only the four individual spectra of {\it XMM-Newton}. The parameters of the broad and narrow \textsc{Gaussian} are given in the second part of Table~\ref{tab:table4}. For constraining the line width of the narrow \textsc{Gaussian} line, we fixed the line width ($\sigma$) to 0.01 keV since it is not accurately resolved by the {\it XMM-Newton} data. The Fe K$\alpha$ line width of the broad component varies from $0.40\pm0.07 \rightarrow 0.33\pm0.05 \rightarrow 0.24\pm0.07 \rightarrow 0.37\pm0.06$ keV. If we calculate the equivalent width of the broad \textsc{Gaussian} component, we see a similarity in 2001 ($EW\sim0.20~{\rm keV}$) and 2016 ($EW\sim0.19~{\rm keV}$) epochs, and 2013 epoch ($EW\sim0.15~\&~0.14 ~{\rm keV}$). The narrow Fe K$\alpha$ line originates from the reflecting clouds located probably in the dusty torus. We estimate the approximate radius ($R_{{\rm Fe~K}\alpha}^{\rm broad/narrow}$) of the broad and narrow Fe K$\alpha$ line assuming the virial motion \citep{Peterson2004,Andonie2022},

\begin{equation}
    R_{{\rm Fe~K}\alpha}^{\rm broad/narrow} = \frac{GM_{\rm BH}}{(\sqrt{3}/2~\nu_{\textsc{fwhm}})^2}~{\rm pc},
\end{equation}
where $G$, $M_{\rm BH}$, and $\nu_{\textsc{fwhm}}$ are the Gravitational constant, the mass of the supermassive black hole, and the full width at half maximum calculated from the best-fitted parameters of \textsc{Gaussian}. It should be noted that we consider that the iron line emission is from the broad line region or from the outer dusty torus, not from the outflow. Otherwise, the virial motion for estimating the radius of the iron emission line would not be valid. We considered the mass of the supermassive black hole to be $5\times10^6$ solar mass. The Gravitational constant value is $4.3\times10^{-3}$ pc~M$_\odot^{-1}$~(km/s)$^2$. We obtain the Fe~K$\alpha$ radius for the broad iron emission line to be $\sim~2.57\times10^{14}$, $\sim~3.67\times10^{14}$, $\sim~6.62\times10^{14}$, and $\sim~3.05\times10^{14}~{\rm cm}$ (or $8.33 \times 10^{-5}~{\rm pc}$, $1.19 \times 10^{-4}~{\rm pc}$, $2.15 \times 10^{-4}~{\rm pc}$, and $9.88 \times 10^{-5}~{\rm pc}$) respectively for the four observations. For the narrow component of the Fe~K$\alpha$ line, the radius is $3.89\times10^{19}~{\rm cm}$. It is to be noted that this is only an approximation as we could not be able to constrain the line width properly.

From the estimated radius of the iron emission line, the $R_{\rm dust}$ and the $R_{\rm BLR}$ (see, Section~\ref{absorption}), we conclude that the emission comes from two different regions of the system. Broad iron line emission comes from very close to the central engine, possibly from the accretion disk, even closer than the BLR, and thus shows a variable nature. The narrow emission line possibly comes from the outer region of the torus and thus shows a constant nature.

\section{Conclusions}
We study the accretion properties of NGC 7314 using {\it XMM-Newton}, {\it NuSTAR}, and {\it RXTE}/PCA data. The {\it XMM-Newton} data covers 15 years (2001 to 2016) with four observations, and {\it NuSTAR} observation was taken in 2016 simultaneous with one of the {\it XMM-Newton} observations. The RXTE/PCA spans from 1999 to 2002. To summarize our findings-
\begin{itemize}
    \item The source shows greater variability in the soft rather than the hard band. The high-energy photons most likely come from the scattering of the more variable soft photons in a hot plasma, located away from the center, producing less variable high-energy photons.
    \item {\it RXTE}/PCA spectral analysis reveals a slow evolution of the accretion properties over time.
    \item The Fe~K$\alpha$ lines come from two different regions. The broad line comes from very close to the SMBH with an approximate radius of $10^{14}$~cm, a high ionization region, most likely from the accretion disk. The narrow component comes from a neutral region, far away from the center, most likely from the molecular region of the dusty torus.
    \item The observed absorption feature could be from the clouds moving around along the line of sight. However, the line-width type absorption indicates to being a high ionizing origin. As the variability of the absorption feature is not so significant, we can assume that these clouds could possibly be located close to BLR ($\sim10^{16}$~cm) in some high ionization region.
    \item The soft excess with a peak energy of around 0.05 keV could be a byproduct of the fewer scattering of the primary photons in the hot plasma that produces high-energy photons.
    Being in the same origin as the less variable high energy photons, we noticed almost no variability for this component during our studied period.
    \item The similar pattern in the spectral properties along with the variability in 2001 and 2016 observations than 2013 observations suggest that NGC 7314 could be a potential candidate for a changing state AGN. To further justify this claim, we propose that continued multi-wavelength monitoring of NGC 7314 is essential. Future observations should focus on detecting any shifts in the spectral state, such as transitions from a Seyfert 1.9 to a Seyfert 1 type or vice versa, which could be accompanied by the appearance or disappearance of broad emission lines or significant changes in the soft X-ray excess. Additionally, long-term monitoring could reveal trends in the variability patterns, which would indicate changes in the inner accretion disk structure.
\end{itemize}

\begin{acknowledgments}
We would like to thank the anonymous referee for their constructive comments and suggestions, which have significantly improved the quality of this paper. D.C. and H.K.C acknowledge the grants NSPO-P-109221 of the Taiwan Space Agency (TASA) and NSTC-112-2112-M-007-053 of the National Science and
Technology Committee of Taiwan. A.J. acknowledges support from the Fondecyt fellowship (Proyecto 3230303)
\end{acknowledgments}

\vspace{5mm}
\facilities{{\it RXTE, XMM-Newton, NuSTAR}}


\software{HeaSoft \url{https://heasarc.gsfc.nasa.gov/docs/software/heasoft/}, {\tt pyXspec-Corner} \url{https://github.com/garciafederico/pyXspecCorner}. 
}

\appendix
\begin{figure}[!ht]
	\includegraphics[width=20cm]{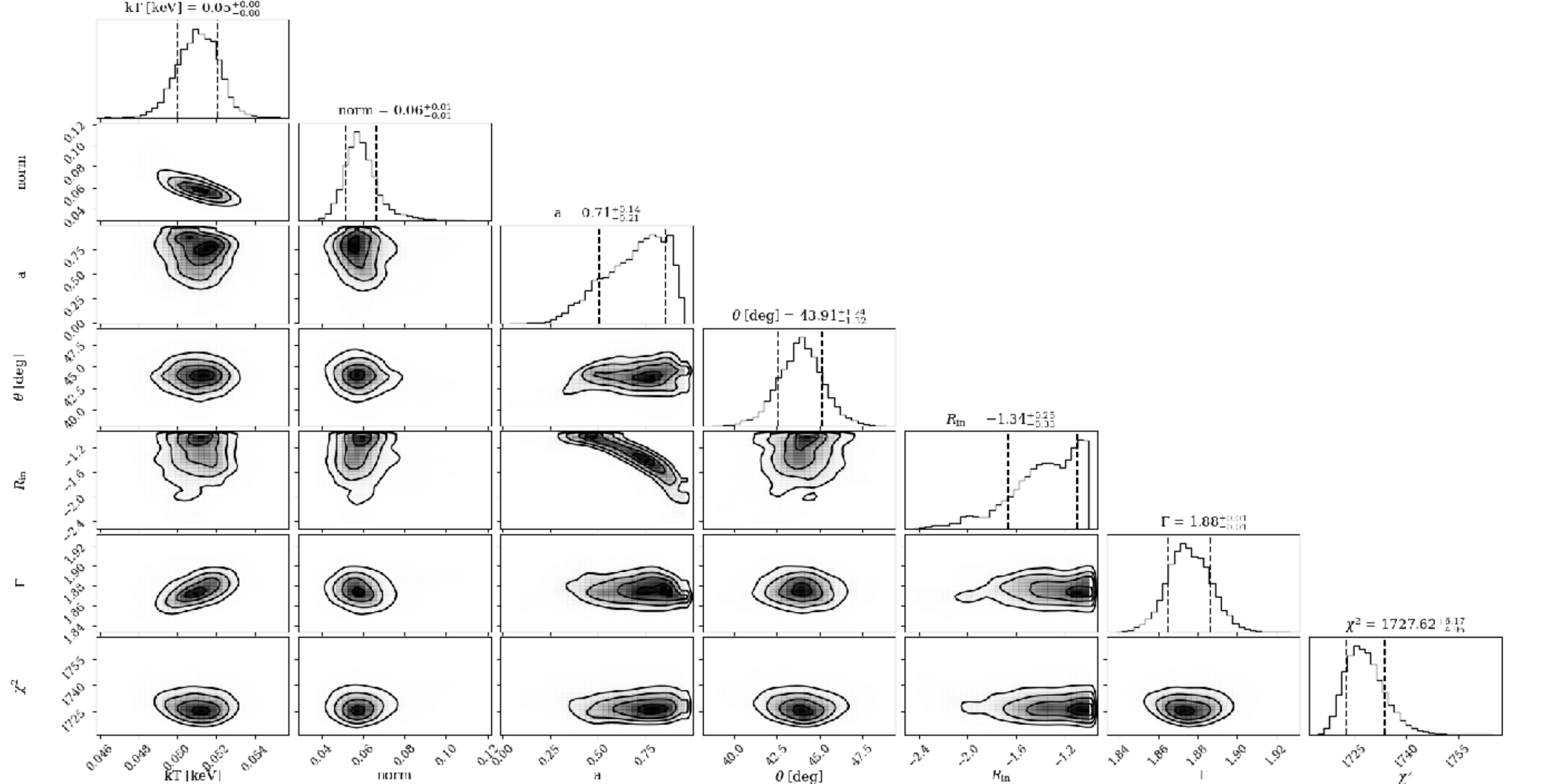}
    \caption{Corner plot of a selection of best-fitting parameters of the \textsc{relxill+Bbody+xillver} model. The contours in the 2D
    histograms show 68, 90, and 95 \% confidence levels. All the values of the best-fitted parameters are listed in Table~\ref{tab:table2}, Model 4.
    }
    \label{fig:corner}
\end{figure}



\bibliography{ngc7314}{}
\bibliographystyle{aasjournal}



\end{document}